\documentclass[twocolumn,showpacs,preprintnumbers,amsmath,amssymb,aps,prb,
superscriptaddress, floatfix ]{revtex4}
\usepackage{graphicx}
\usepackage[latin1]{inputenc}

\newcommand{\be}{\begin{equation}}
\newcommand{\ee}{\end{equation}}
\newcommand{\bea}{\begin{eqnarray}}
\newcommand{\eea}{\end{eqnarray}}

\newcommand{\ri}{\mbox{i}}
\newcommand{\re}{\mbox{e}}

\begin{document}
\title{Fractionalization in Three-Components Fermionic Atomic Gases in a One-Dimensional Optical Lattice}
\author{ Patrick Azaria  }
\affiliation{Laboratoire de Physique Th\'eorique de la Mati\`ere Condens\'ee, Universit\'e Pierre et Marie Curie and CNRS (UMR 7600), 4 Place Jussieu, 75256 Paris Cedex 05, France}
\date{\today}
\pacs{
{03.75.Mm},  
{71.10.Pm},  
{71.10.Fd},  
}
\begin{abstract}
 We study a three-components fermionic gas loaded in a one-dimensional optical trap at half-filling. We find 
 that the system is fully gapped and may order into 8 possible phases: four 2$k_F$ atomic density wave and 
 spin-Peierls phases with all possible relative $\pi$ phases shifts between the three species. We find that trionic excitations 
 are unstable toward the decay into pairs of kinks carrying a {\it fractional} number, $Q=3/2$, of atoms. These sesquions
 eventually condense upon small doping and are described by a Luttinger liquid. We finally discuss the phase diagram
 of a three component mixture made of three hyperfine level of $^6$Li as a function of magnetic field.
\end{abstract}
\maketitle
\sloppy
A problem analogue to  quark  confinement in particle physics has been recently addressed 
in systems with multi-components fermionic  atoms loaded in an optical 
 lattice\cite{phle,Wu,miyake,rapp,capponi,Liu2008,Guan2008,roux2008,dukelsky2008, ottenstein}.
These studies strongly support the formation of  ``baryonic" states made of bound states 
of $n>2$ atoms. In one dimension, for example, trionic states, 
made of bound state of three atoms, were predicted to be stable at generic densities and sufficiently
 low temperatures, typically of the order $\sim 30-100$nK\cite{azaria2}. 
This result   opens the exciting possibility to probe in a new context  and in future experiments a simplified version
of  the  ``quark" confinement phenomenon in quantum chromodynamics.
In all these  previous studies  the attention has been drawn on the formation 
of  baryonic (or molecular) states that contains an {\it  integer} number
of atoms. Here we shall focus on  the intriguing  situation where
 the low-energy elementary  excitations  carry  a {\it fractional}  number of atoms.
Although it may appears counter-intuitive,  fractionalization of quantum numbers 
is a well established phenomenon in  condensed matter physics.  Celebrated examples  
are fractionally charged excitations in the quantum Hall state\cite{laughlin}   and in
quasi-one-dimensional polymers\cite{su}.  In this work we shall present strong arguments 
that fractionalization can also occur in ultra-cold atomic physics. 
We shall give  evidences that at densities
close to  half-filling  a three-components fermionic mixture loaded in a one-dimensional
optical trap may support low-energy excitations  carrying a fractional number, $Q=3/2$, of atoms, the sesquions. 

When loaded in a one-dimensional optical lattice of wavelength $\lambda$,
 a three-components mixture is well described,
away from resonance, by  a Hubbard-type  hamiltonian of the form\cite{Jaksch}:
\be
{\cal H} = -t \sum_{i,a} \left[ c^{\dagger}_{i,a} c^{ }_{i+1,a} +  {\rm H.C} \right] +
\sum_{i, a<b} U_{ab} \;  \rho_{i,a}\rho_{i,b} 
\label{hamiltonianani}
\ee
where  $c^{\dagger}_{i,a}$ is the creation operator for a fermionic atom of species  $a=(1, 2, 3)$, at site $i$, and
  $\rho_{i,a} = c^{\dagger}_{i,a}c_{i,a}$ is the local density of the atomic species $a$. The parameters $t$ and the
 couplings $U_{ab}$ can be expressed in term of the recoil energy, the laser intensity and wavelength as well as 
 the s-wave scattering lengths $s_{ab}(B)$ between the species. For generic external magnetic fields $B$, the $s_{ab}$ are
 in general different and so are the couplings $U_{ab}$. Thus the physical
 symmetry group  of  (\ref{hamiltonianani}) is  $U(1)^3$ corresponding to the conservation of the number
 of atoms of each species. Such a small symmetry, which   is an essential feature of  atomic mixtures, make the elucidation of 
 the physics  associated with (\ref{hamiltonianani}) a difficult task. However, as we shall see, much can be said in the weak coupling, low-energy, limit.
 The physics described by (\ref{hamiltonianani}) strongly depends on the  density of atoms $\bar \rho= \bar \rho_1 =\bar \rho_2 =\bar \rho_3$.
Away from half-filling, i.e. when  $\bar \rho\neq 1/2$, it was shown in Ref.~\onlinecite{azaria2} that for generic couplings the dominant fluctuations 
consists into massless $2k_F$ Atomic Density Waves (ADW) and massless trionic excitations carrying total atomic number $Q=3$.
At half filling, when $\bar \rho = 1/2$, the physics is, as we shall see, radically different.

{\it Effective Low Energy Hamiltonian}.
The low energy effective theory associated with the Hubbard Hamiltonian (\ref{hamiltonianani}) can 
be derived, as usual, from the linearization  at the two Fermi points $\pm$k$_F$ of the 
dispersion relation of free three-component fermions:
\be
c_{i,a} \sim \Psi_{aR} \; \re^{i k_F x} + \Psi_{aL} \; \re^{-i k_F x} \; \; a=(1,2, 3),
\ee
 where $x = i a_0$, $a_0=\lambda/2$ is the lattice spacing, $\lambda$ the laser wavelength, and k$_F$ = $\pi{\bar \rho}/a_0$ 
  is the Fermi wave-vector. Finally  $\bar \rho=1/2$ is the density per species.   In the weak coupling limit $|U_{ab}|/t << 1$ 
  the effective hamiltonian associated with (\ref{hamiltonianani}) is found to be:
 \be
 {\cal H} = {\cal H}_0 + {\cal H}_{\rm I}
 \label{h}
 \ee
 with
 \bea
 {\cal H}_0= - \ri v_F \sum_a (\Psi^{\dagger}_{aR}\partial_x \Psi^{}_{aR} - \Psi^{\dagger}_{aL}\partial_x \Psi^{}_{aL})
 \label{h0}
 \eea
 and
 \bea
 {\cal H}_{\rm I} &=& \sum_{a<b} \left( \mu_{ab}\;  h_{aR} h_{bL} + \lambda^{-}_{ab} \; I^{\dagger}_{abR}I^{}_{abL} +
 \lambda^{+}_{ab} \; J^{\dagger}_{abR}J^{}_{abL}\right) 
 \nonumber \\
 &+&  
 (R \leftrightarrow L) 
 \label{hint}
 \eea
where
\bea
h_{aR(L)} = \Psi^{\dagger}_{aR(L)} \Psi^{}_{aR(L)},
I^{\dagger}_{abR(L)} =  \Psi^{\dagger}_{aR(L)} \Psi^{}_{bR(L)},
\eea
and 
\be
J^{\dagger}_{abR(L)} = \Psi^{\dagger}_{aR(L)} \Psi^{\dagger}_{bR(L)}.
\ee
 The fermi velocity is given by $v_F= 2t$, $\mu_{ab}= U_{ab}$, $\lambda^{\pm}_{ab} = \pm U_{ab}$ and we have omitted
 a term $ \sum_{a<b}({h}_{aR}{h}_{bR} + {h}_{aL}{h}_{bL})$ that account for a non-uniform velocity renormalization. 
 In the absence of the $\lambda^+$-term  we recover the hamiltonian studied in Ref.~\onlinecite{azaria2} in the $\bar \rho \neq 1/2$ case
 which  involves   the $9$ $U(3)|_{R(L)}$ currents ${\cal J}_{\parallel R(L)} \equiv (h_{aR(L)}, I^{\dagger}_{abR(L)}, I^{}_{abR(L)})$.
 The $\lambda^+$-term is a 4$k_F$ contribution of the density-density interaction among the species and is present 
 only at half-filling. It involves the $6$ currents ${\cal J}_{\perp R(L)} \equiv (J^{\dagger}_{abR(L)}, J^{}_{abR(L)})$ that create or destroy pairs of
 atoms. Together with the $U(3)|_{R(L)}$ currents they  generate the $15$  $SO(6)|_{R(L)}$ currents
 ${\cal J}_{R(L)} = {\cal J}_{\parallel R(L)} \oplus {\cal J}_{\perp R(L)}$.  The non interacting part  of (\ref{h}), $ {\cal H}_0$,
  is that of relativistic free fermions and  has the maximally available symmetry $SO(6)_R \otimes SO(6)_L$ generated by 
 the  ${\cal J}_{R(L)}$'s. The interaction hamiltonian breaks the later  symmetry down to $U(1)^3_R \otimes U(1)^3_L|_{\rm diag}$ 
 corresponding to the conservation of the number of atoms of each species $a$. Though the lattice hamiltonian (\ref{hamiltonianani})
 depends on three couplings $U_{ab}$ the hamiltonian (\ref{h}) is the most general
 hamiltonian for the three species problem with an $U(1)^3$ symmetry and one has to consider the role  
 of the  $15$ couplings ($\lambda^{\pm}_{ab}, \mu_{ab}$) that
 encode all possible competing orders.   Which one is likely to be stabilized in  the low energy limit depends on the 
 asymptotic behavior  of the Renormalization Group (RG) flow. 

 \begin{figure}[t]
 \includegraphics[width=6.5cm,clip]{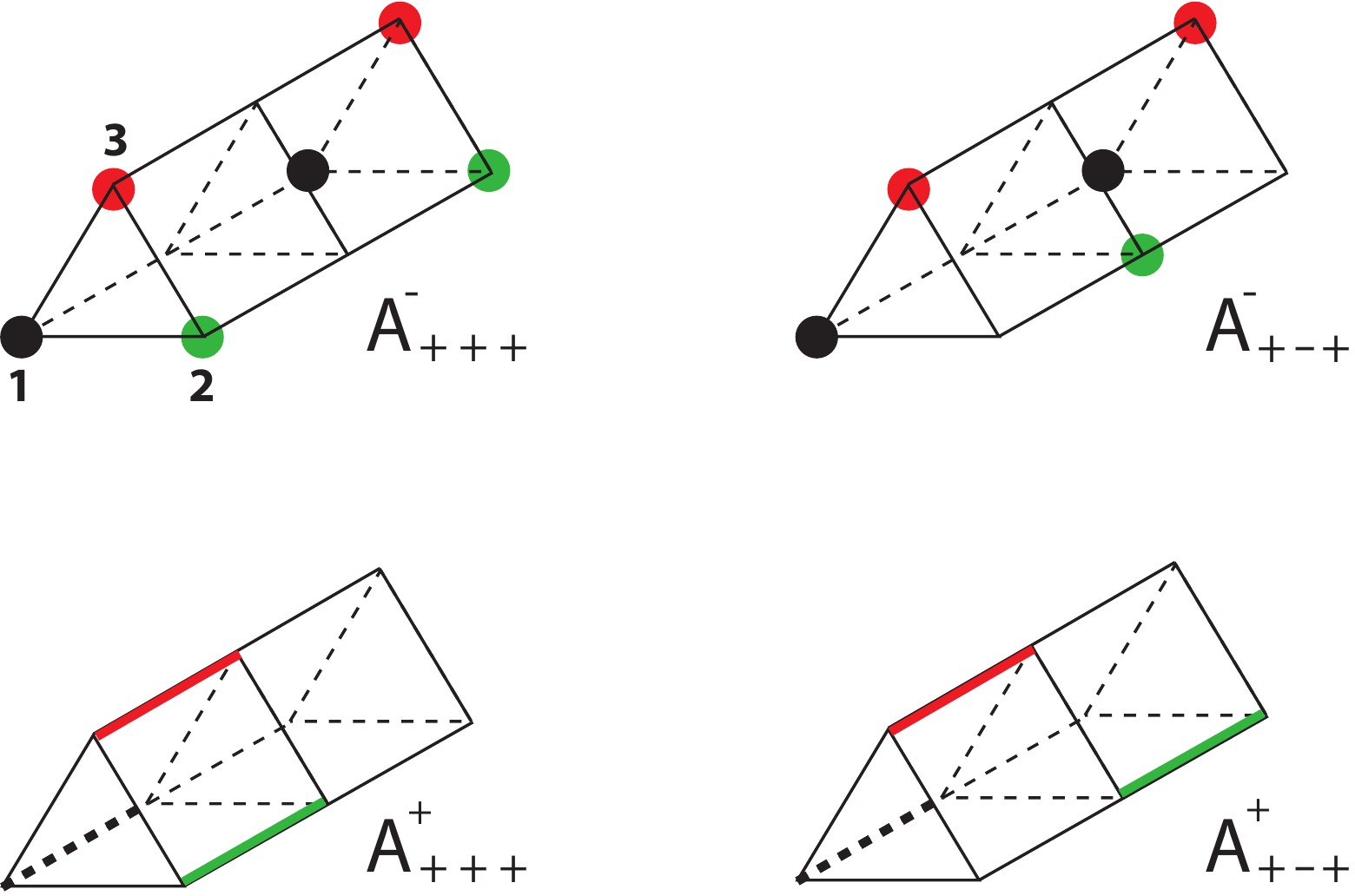}
 \caption{Generalized  atomic density wave  and spin-Peierls phases. Atomic species are 
 labelled $1,2$ and $3$. Among the $8$ possible phases ${ A}^{\pm}_{+++}$, ${ A}^{\pm}_{++-}$, ${ A}^{\pm}_{+-+}$,
 ${ A}^{\pm}_{-++}$ four are represented. In the  ${ A}^{\pm}_{+-+}$ phase the species $1$ and $3$ are in phase and out of phase with
 the species $2$. }
 \label{fig1}
 \end{figure}

 {\it Renormalization Group Analysis}.
We have obtained the one-loop  RG equations associated with (\ref{h}). 
They will be given  elsewhere\cite{azaria3} and we shall only  present in the following
our  results. Due of the lack of symmetry in the problem, and consequently of the relatively  large number 
 of independent couplings $(\lambda^{\pm}_{ab}, \mu_{ab})$,
it may seems an akward task to draw any general conclusions on the phase diagram associated
with (\ref{h}). Fortunately, it has been recognized\cite{lin} that due to the importance of strong quantum fluctuations
in one dimensional systems the symmetry at level of the lattice spacing $a_0$ is likely to be enlarged at low energies which
thus considerably simplify  the problem. Phrased in the  RG language
such a  Dynamically Symmetry Enlargement (DSE) correspond to a situation where the
hamiltonian (\ref{h}) is attracted under the RG flow toward an effective hamiltonian
$H^*$  with a higher symmetry. As shown in Ref.~\onlinecite{boulat}  the possible DSE fixed points $H^*$ 
 depend only on the symmetry breaking pattern described by  (\ref{h}). In the present case we find that
 for {\it generic} initial conditions  of the RG flow, $(\lambda^{\pm}_{ab}, \mu_{ab})$, the low-energy 
 physics associated with (\ref{h}) is described by one of the  fixed points hamiltonians:
\bea
{\cal H}^{\pm}_{ \epsilon_1 \epsilon_2\epsilon_3}&=& - \ri v \sum_a (\Psi^{\dagger}_{aR}\partial_x \Psi^{}_{aR} - \Psi^{\dagger}_{aL}\partial_x \Psi^{}_{aL})
 \nonumber \\
&\pm&  G \; \left( \sum_a \epsilon_a (\Psi^{\dagger}_{aR} \Psi^{}_{aL} \mp \Psi^{\dagger}_{aL} \Psi^{}_{aR} )\right)^2
\label{so2nepsilon}
\eea
 where  $v$ is a renormalized velocity, $\epsilon_a = \pm1$, and  $G$ is some positive coupling.
 As  (\ref{so2nepsilon}) is invariant under the simultaneous change $\epsilon_a \rightarrow -\epsilon_a$,
 there are  $8$ independent fixed points, with $(\epsilon_1 \epsilon_2\epsilon_3)= (+++), (-++), (+-+)$ and $(++-)$, 
that  describe  phases, labelled ${ A}^{\pm}_{\epsilon_1 \epsilon_2\epsilon_3}$, 
 with qualitativally different physical properties.  The phases   ${ A}^{+}_{\epsilon_1 \epsilon_2\epsilon_3}$ and ${ A}^{-}_{\epsilon_1 \epsilon_2\epsilon_3}$, 
 are generalized $2k_F$, Spin-Peierls (SP) and Atomic Density Wave (ADW) phases. The $\epsilon_a$ account for all possible  relative
  $\pi$-phase shifts between the species $a$. A pictorial representation of the ground states  is presented 
  in Fig.\ref{fig1}. The corresponding lattice order parameters can be readily obtained from the structure of the interacting part of (\ref{so2nepsilon}) and are
given by:
\bea
 {\cal O}^{+}_{\epsilon_1 \epsilon_2 \epsilon_3} &=&  \sum_{a} \frac{(-1)^i}{2}  \epsilon_a ( c^{\dagger}_{ai} c^{}_{a i + 1 } + {\rm h.c}) \\ \nonumber
 {\cal O}^{-}_{\epsilon_1 \epsilon_2 \epsilon_3} &=&   \sum_{a} \frac{(-1)^i}{2} \epsilon_a ( c^{\dagger}_{ai} c^{}_{a i  } + {\rm h.c}).
 \label{Opmepsilon}
  \eea
  In each phase the ground state is doubly degenerated and when $< {\cal O}^{\pm}_{\epsilon_1 \epsilon_2 \epsilon_3}> \neq 0$
  there is spontaneous symmetry breaking of translational invariance by one lattice site. As a consequence we expect
  kinks (or solitonic)  excitations that interpolate between the two ground states to be present in the spectrum. As we shall
  see these have fractional quantum numbers.
 
 {\it Spectrum and Fractionalization}.
Remarkably enough, the different hamiltonians (\ref{so2nepsilon}) can be brought to the same form by mean
of duality transformations\cite{boulat}:
 \be
{\cal H}^{\pm}_{\epsilon_1 \epsilon_2\epsilon_3}( G, \Psi) ={\cal H}^{+}_{+++}(G, \omega^{\pm}_{\epsilon_1 \epsilon_2\epsilon_3}(\Psi)),
\label{hduality}
\ee
where the duality transformations   $\omega$ act only on  the right-moving fermions as:
\be
  \omega^{\pm}_{\epsilon_1 \epsilon_2\epsilon_3}(\Psi^{}_{aR})={ \displaystyle \re^{{i\frac{\pi}{4}(1\mp1)}}  \;   \epsilon_a \Psi^{}_{aR}}.
  \label{dualities}
\ee
We therefore find that the elucidation of the low-energy physics described by the fixed points hamiltonians 
${\cal H}^{\pm}_{\epsilon_1 \epsilon_2\epsilon_3}$ stem from the knowledge of those of   $H^{+}_{+++}$.
The latter hamiltonian has an enlarged $SO(6)$ symmetry generated by the
 ${\cal J}^A = \int dx ({\cal J}^A_R + {\cal J}^A_L)$, $A=(1,..., 15)$, 
 and is that of the $SO(6)$ Gross-Neveu (GN) model.
 The other fixed points hamiltonians (\ref{so2nepsilon}) has a dual extended symmetry 
$ {\tilde SO(6)} $ generated by the dual currents
${\tilde{ \cal J}}^A = \int dx ( \omega^{\pm}_{\epsilon_1 \epsilon_2\epsilon_3}({\cal J}^A_R ) + {\cal J}^A_L)$.
 Fortunately, the $SO(6)$ GN model is integrable\cite{andrei} so that  its spectrum  is exactly known and by duality  the  one 
 in the other phases ${ A}^{\pm}_{\epsilon_1 \epsilon_2\epsilon_3}$. In all cases it consists into two set of 
 four kinks, $S^{\pm}_{\alpha=0,1,2,3}$,  of mass $m_S \sim t\re^{-t/U}$  ($U$ being a characteristic energy scale), belonging to 
 the two spinorial representations of $SO(6)$ (or $ {\tilde SO(6)} $) and  six  real (Majorana) fermions $\xi_{\beta=1,..., 6}$ of mass $m_F=\sqrt 2 \; m_S$ 
 transforming according to  the vectorial representation of $SO(6)$  (or $ {\tilde SO(6)} $).
Though their wave functions are different in all the phases ${ A}^{\pm}_{\epsilon_1 \epsilon_2\epsilon_3}$, these particles
 are described by the same quantum numbers since the duality transformations  (\ref{dualities})  do not affect
 the $3$  conserved charges (or Cartan generators):
\be
q_a = \int dx \; (\Psi^{\dagger}_{aR}\Psi^{}_{aR} + \Psi^{\dagger}_{aL}\Psi^{}_{aL}), \; a=(1, 2, 3).
\label{qa}
\ee
These are nothing but the total number of atoms of  a given species  $a$ and  the particles of the spectrum
 are labelled by the set of quantum numbers $(q_1, q_2, q_3)$.  The kinks quantum numbers are {\it fractional} :
 $S^{\pm}_{0}=\pm1/2(1,1,1), S^{\pm}_{1}=\pm1/2(1,-1,-1), S^{\pm}_{2}=\pm1/2(-1,1,-1)$ and $S^{\pm}_{3}=\pm1/2(-1,-1,1)$.
The fermions are bound states of two kinks and have the same quantum numbers as the original lattice fermions: $\xi_{\beta}= (\pm1, 0, 0), (0, \pm1, 0)$ and  $(0, 0, \pm1)$.
There are no other stable particles. In particular there are no stable trions in contrast with what happens at incommensurate  fillings.
Trionic excitations, $T^{\dagger} = c^{\dagger}_1  c^{\dagger}_2  c^{\dagger}_3$,  have total atomic number $Q=3$, where
\be
Q= q_1 + q_2 + q_3,
\label{Q}
\ee
and are unstable toward the decay into elementary kinks or fermions. A  trion has quantum numbers $(1,1,1)$ 
and the  most energetically favorable process is $T^{\dagger} \rightarrow S^{+}_{0} S^{+}_{0}$ 
so that  one may think of the kink $S^{+}_{0}$ as  ``half" a trion. As it has  total atomic number $Q=3/2$ one may call it a sesquion.
The existence of these fractional kinks as the lowest energy excitations in generic three-components 
ultra-cold atomic systems is an unexpected and non-trivial finding and constitute one of the main results
of the present work. It is therefore worth discussing the stability of  the above excitations.
Indeed, as noticed in both Refs.~(\onlinecite{lin,boulat}), the DSE
is only approximate and we expect residual  symmetry breaking operators
to survive even in the low-energy limit.  As the $SO(6)$ GN particles are  labeled
by the conserved quantum numbers (\ref{qa})  associated with the $U(1)^3$ symmetry
of the problem, small residual anisotropy will result only into a small splitting of the particle
spectrum.  For large enough anisotropy and/or strong couplings   it may eventually happens that  the above
description of the spectrum  breaks down. We expect however that 
the DSE description of the hamiltonian (\ref{hamiltonianani}) (and hence the stability of the sesquions)
 holds in a relativly  large portion of the phase diagram. Indeed the accuracy of the DSE description  has been checked 
 numerically in the different context of  the $SU(4)$
Hubbard model at half-filling where the adiabatic continuity of the $SO(8)$ GN  spectrum
in this case  has been explicitly observed for small enough interactions\cite{assaraf}.

{\it Doping}. To model small doping we consider adding a chemical potential term $H_Q = - \mu Q$ (we consider  hole doping with $\mu > 0$). As $Q$ is 
invariant under the duality transformation (\ref{dualities}) it is sufficient to consider doping the $SO(6)$
Gross-Neveu model. The chemical potential term breaks the $SO(6)$ symmetry but since  $[H^{+}_{+++} , Q] = 0$ doping 
does not spoil integrability and the following picture emerges.
At non zero $\mu$, the particle spectrum is splitted according to the values of $Q$. A particle with mass $m$ and atomic number $Q$
will lower its energy to $m - \mu Q$. When  its  energy becomes negative the ground state start to fill 
with these particles. When $\mu > 2m_S/3$ the first particle that start filling the ground state is the sesquion $S^{+}_{0}$ of mass $m_S$ and $Q=3/2$.
 As $\mu$ is increased further other particles would like to enter the ground state like other members of the 
kinks multiplets with $Q=1/2$ or the fermions with $Q=1$. However the increase of the chemical potential is counteracted
by the repulsion  felt by the kinks and the fermions to the sesquions\cite{evans, saleur}. 
As a result for $\mu > 2m_S/3$ the ground state is  only filled by sesquions which become
massless excitations. The effective theory describing these massless fractional $Q=3/2$ excitations 
is a Luttinger liquid with a stiffness $K$. At these dopings the kinks $S^{\pm}_{\alpha=1,2,3}$ and the fermions
$\xi_{\beta}$ remain massive, with renormalized masses.  Both the renormalized masses and the stiffness $K$
could be in principle  computed from the Bethe ansatz solution in a similar way as done in Ref.~\onlinecite{saleur}.
At large doping, i.e. when $\mu >> m_S$, the 4$k_F$ term $\lambda^+$ in (\ref{hint}) decouples and one
recovers the physic described above for the generic filling case with massless $Q=3$ trionic excitations.
We may therefore expect that below some critical value of the density $ \bar \rho < {\bar \rho}_c$ sesquions get confined
into trions. The elucidation of the nature as well as the location of such a confinement/deconfinement transition goes 
beyond the scope of this work and will studied elsewhere\cite{azaria3}.
 \begin{figure}[t]
 \includegraphics [width=9cm,clip]{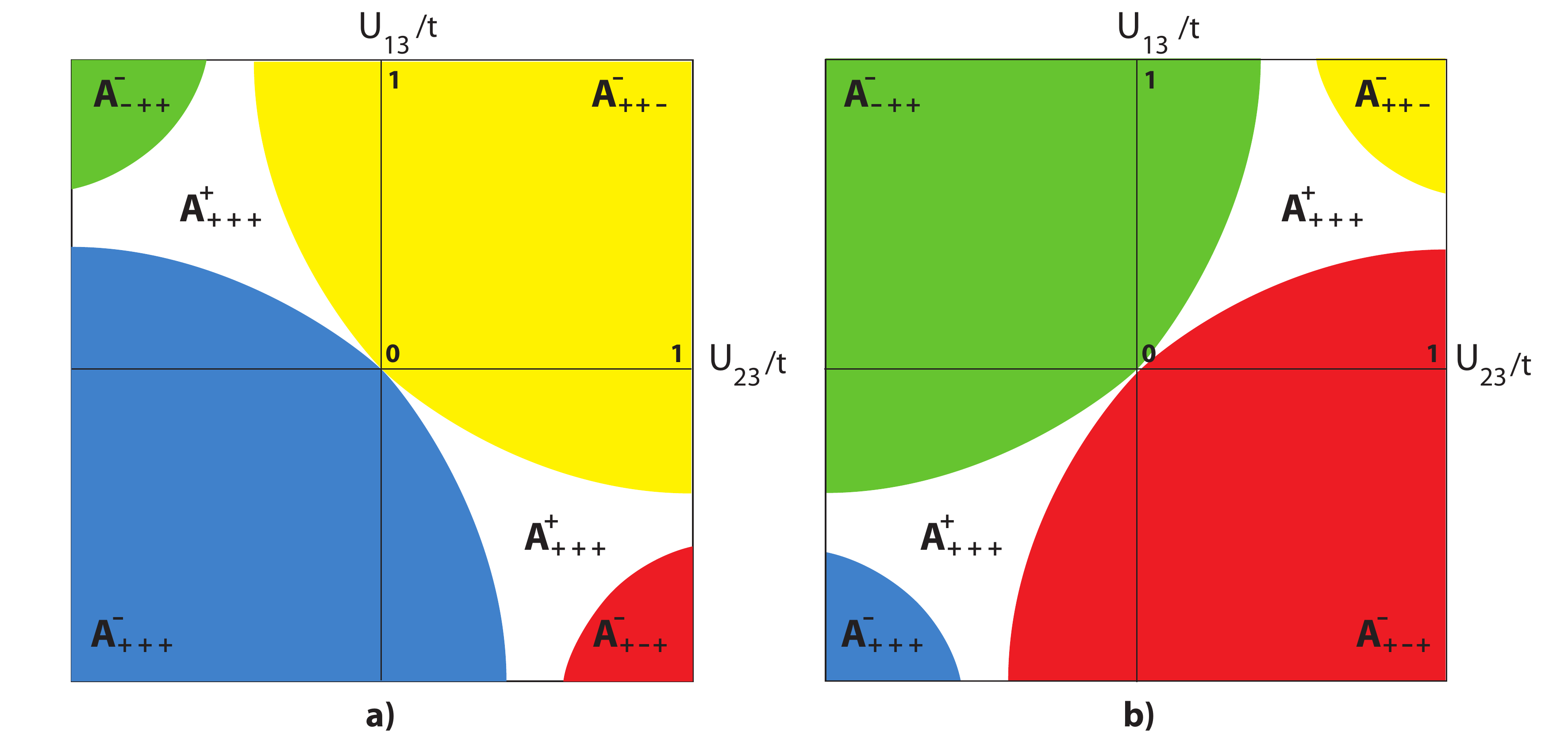}
 \caption{2D projections of the weak coupling phase diagram  of the anisotropic Hubbard model for fixed values of $U_{12}/t$:  a) $U_{12}/t = - 0.5$ and
  b) $U_{12}/t = 0.5$. The SP phase ${ A}^{+}_{+++}$ occurs in the frustrated region (i.e. when $(\Pi_{a<b} \; U_{ab} ) > 0$) around the isotropic
  points $|U_{ab}| = U$.}
 \label{fig2}
 \end{figure}

{\it Three-Species Problem and Experiments}. 
 The phase diagram  in the three-dimensional space $(U_{12}/t, U_{23}/t,U_{31}/t)$ is rich and complex, revealing the delicate balance
 between the different competing orders. We find that among the $8$ possible phases only $5$  are stabilized:
 the four ADW phases ${ A}^{-}_{\epsilon_1 \epsilon_2\epsilon_3}$ and the uniform SP phase ${ A}^{+}_{+++}$. We show 
 in Fig.2. two projections of the phase diagram in the $(U_{23}/t,U_{31}/t)$ plane for typical  value of $U_{12}/t = \pm 0.5$.
 Though it is difficult to draw any general quantitative picture  of the phase diagram we observe that: i) in the unfrustrated 
 regions, where $(\Pi_{a<b} \; U_{ab} ) <  0$, the ADW phases that are stabilized are the one
 that minimize the  density-density potential in (\ref{hamiltonianani}) and ii) that in the frustrated regions, when $(\Pi_{a<b} \; U_{ab} ) > 0$,
the  kinetic energy term play an important role  when all couplings  are of the same order of magnitude. In particular in the vicinity
of the  isotropic rays $|U_{ab}| = U$ a  uniform SP phase  ${ A}^{+}_{+++}$ is likely to be stabilized.  Eventually the above SP phase is
destabilized in favor of  various ADW phases for sufficiently large  anisotropies.
 In experiments once the optical lattice parameters  and the density  are  fixed the only control 
 parameter is the external magnetic field $B$.  The phase diagram as a function of the magnetic
 field $B$ is a  line in the three-dimensional space $(U_{12}(B)/t, U_{23}(B)/t,U_{31}(B)/t)$ which dependence 
 on  $B$  essentially depends on the mixture through the s-wave scattering lengths.
 Taking for example~\cite{ottenstein} a mixture made of a 
balanced population of  three  hyperfine states  of $^6$Li
 atoms, $ |F,m_F\rangle = |1\rangle=|1/2,1/2\rangle, |2\rangle=|1/2,-1/2\rangle$, and $|3\rangle= |3/2,-3/2\rangle$, 
with typical optical lattice parameters~\cite{azaria2} and a laser wavelength $\lambda = 1\mu$m, we find  a weak coupling
regime where a non trivial SP phase may be observed. At half-filling
using our one loop RG equations the following phase diagram as a function of the magnetic field emerges. For small  fields
$B < B_{c1}$, a uniform ADW phase ${ A}^{-}_{+++}$ is stabilized while at larger fields $B  > B_{c2}$
an ADW phase ${ A}^{-}_{-++}$, where the species labeled $1$ is in phase opposition with species $2$ and $3$, shows off.
This match the result of Ref.~\onlinecite{azaria2} for $\bar \rho \neq 1/2$ where at these filings 
ADW phases of the same type with quasi-long range order were predicted. The essential difference with the above case is that when 
$\bar \rho = 1/2$ an intermediate uniform SP phase ${ A}^{+}_{+++}$ is locked in the region $ B_{c2} < B < B_{c1}$.
Within the one loop accuracy we find $B_{c2} \sim 560$G and $B_{c1} \sim 540$G. This is  an interesting result
from the experimental point of view since in both ${ A}^{+}_{+++}$ and ${ A}^{-}_{-++}$ phases we expect that 
effect of the three-body losses\cite{azaria2}  will be considerably reduced.  The knowledge of actual values of the binding energies
of the sesquions and hence of the temperature scale below which these phases could be stabilized call for an alternative
approach such like numerical calculations\cite{azaria3}.
To summarize we have shown that in the vicinity of half-filling fractional excitations
carrying $Q=3/2$ atoms, the sesquions, are the relevant low energy excitations
in a generic  three-components Fermi mixture. These sesquions are likely to  get confined into $Q=3$ trionic
excitations when one moves sufficiently far away from half-filling. We therefore expect that both
the confined (trionic) and unconfined (sesquionic) phases could be probed in future experiments.

We thank E. Boulat, S. Capponi, V. Dubois, P. Lecheminant and M. Najac for discussions and encouragements.


\begin{thebibliography}{99}
\bibitem{phle}
P. Lecheminant, E. Boulat, and  P. Azaria,
Phys. Rev. Lett. {\bf 95}, 240402 (2005). 
\bibitem{Wu}
C. J. Wu, Phys. Rev. Lett. {\bf 95}, 266404 (2005).
\bibitem{miyake}
H. Kamei and K. Miyake,
J. Phys. Soc. Jpn. {\bf 74}, 1911 (2005).
\bibitem{rapp}
A. Rapp {\it et al.}, 
Phys. Rev. Lett. {\bf 98}, 160405 (2007);
A. Rapp, W. Hofstetter, and G. Zar\'and,
Phys. Rev. B {\bf 77}, 144520 (2008).
\bibitem{capponi} 	
S. Capponi {\it et al.}, 
Phys. Rev. A {\bf 77}, 013624 (2008).
\bibitem{Liu2008}
X.-J. Liu, H.~Hu, and P. D. Drummond, Phys. Rev. A {\bf 77}, 013622 (2008).
\bibitem{Guan2008}
X. W. Guan {\it et al.}, 
Phys. Rev. Lett. {\bf 100}, 200401 (2008).
\bibitem{roux2008}
G. Roux {\it et al.}, Eur. Phys. J. {\bf 68}, 293 (2009)
\bibitem{dukelsky2008}
R. A. Molina {\it et al.}
arXiv: 0807.1886.
\bibitem{ottenstein}
T. B. Ottenstein {\it et al.}, Phys. Rev. Lett. {\bf 101}, 203202 (2008).
\bibitem{azaria2}P. Azaria, S. Capponi and P. Lecheminant, Phys. Rev. A {\bf 80}, 041604(R) (2009).
\bibitem{laughlin} 
R.B. Laughlin,  Phys. Rev. Lett. {\bf 50}, 1395 (1982).
\bibitem{su} W.P. Su, J.R. Schrieffer, and A.J. Heeger, Phys. Rev. Lett. 42, 1698 (1979), 
R. Jackiw and C. Rebby, Phys. Rev. D 13, 3398 (1976)
\bibitem{Jaksch} 
D. Jaksch and P. Zoller, Ann. Phys. (N.Y.) {\bf 315}, 52 (2005).
\bibitem{andrei} 
N. Andrei and J. H.~Lowenstein, Phys. Lett. B {\bf 90}, 106 (1980).
\bibitem{lin} H.H. Lin, L. Balents and M. Fisher, Phys. Rev. B {\bf 58}, 1794 (1998).
\bibitem{boulat} E. Boulat, P. Azaria and P. Lecheminant, Nucl. Phys. B {\bf 822}, 367 (2009).
\bibitem{assaraf} R. Assaraf {\it et al}, Phys. Rev. Lett. {\bf 93}, 016407 (2004)
\bibitem{saleur} R. Konik {\it et al}, Phys. Rev. Lett. 42, 1698 (1998)
\bibitem{evans} J. Evans and T. Hollowood, Nucl. Phys. Proc. Suppl. 45A, 130 (1996). 
\bibitem{azaria3} P. Azaria, S. Capponi and H. Nonne, {\it in preparation}.
\end{thebibliography}
\end{document}